\documentclass[amssymb,aps,twocolumn,showpacs]{revtex4}
\usepackage[]{graphicx}

\def\ket#1{| #1 \rangle}

\def\kb#1#2{| #1 \rangle\!\langle #2 |}
\def\II{1\!\mathrm{l}}

\begin{document}

\title{Testing integrability with a single bit of quantum information}
\author{David Poulin$^{1,2}$, Raymond Laflamme$^{1,2}$, G.J. Milburn$^3$, and  Juan Pablo Paz$^{4,5}$}
\affiliation{
$^1$ Perimeter Institute for Theoretical Physics, 35 King Street N.,
Waterloo, ON, N2J 2W9, Canada. \\
$^2$Institute for Quantum Computing, University of Waterloo, Waterloo, ON, N2L 3G1, Canada.\\
$^3$ Centre for Quantum Computer Technology, School of Physical Science,
The University of Queensland, QLD 4072 Australia. \\
$^4$ Departamento de F\'{\i}sica ``J.J. Giambiagi", FCEN, UBA, Pabell  on 1, Ciudad Universitaria, 1428 Buenos Aires, Argentina. \\
$^5$ Theory Division, MS B213, Los Alamos National Laboratory, Los Alamos, NM 87545
}
\date{\today}

\begin{abstract} 
  We show that deterministic quantum computing with a single bit
  (DQC1) can determine whether the classical limit of a quantum system
  is chaotic or integrable using $O(N)$ physical resources,
  where $N$ is the dimension of the Hilbert space of the system under
  study. This is a square root improvement over all known classical
  procedures. Our study relies strictly on the random matrix
  conjecture. We also present numerical results for the nonlinear
  kicked top.
\end{abstract}


\maketitle

\section{Overview} \label{intro}

After an initial triumph at solving mathematical problems (see
\cite{NC2000} for an overview), a large fraction of the researches in
the field of quantum information processing has shifted to its
original motivation: the simulation of quantum systems
\cite{Feynman1982}. It is now well established
\cite{Lloyd1996,OGKL2001,SOGKL2002} that the evolution produced by
certain classes of Hamiltonians can be simulated efficiently on a
universal quantum processor. However, extracting useful information
from the physical simulation is a problem who's complexity has been
underestimated. Indeed, the ability to simulate the dynamics of a
system does not grant one with the ability to evaluate efficiently all
physical quantities of interest. These quantities (e.g. spectral
properties) are usually measured experimentally on a (exponentially)
large number of physical systems --- macroscopic samples. A direct
quantum simulation, on the other hand, can only reproduce the
statistical output of a single quantum system which yields drastically
less information than what is learned from costly classical
simulations.  Thus, it is not clear at this point whether quantum
simulators can always outperform their classical analogues.

Some of these spectral properties play a central role in the study of
quantized chaotic systems. One particular question of interest is
whether the classical limit of a quantum system exhibits regular or
chaotic motion. It has became widely accepted (see
\cite{Haake2001,Stockmann1999} and references therein) that the answer
to this question is hidden in some spectral properties of the system,
which can be reproduced by those of canonical random matrices with the
appropriate symmetries. Given a description of the Hamiltonian of the
system, the
best known algorithms evaluating these ``signatures of chaos'' require
classical computing resources which grow at least as fast as $N^2$, the
square of the dimension of the Hilbert space of the system under study. 
Indeed, a close inspection of these algorithms show that they require either
matrix multiplication, diagonalization, or evaluation of a determinant \cite{Haake2001}. 
Since such a
growth is intractable on any conventional computer (remember that $N$
grows exponentially with the size of the physical system), it is quite
natural to try to tackle this problem with a quantum computer. In
recent years, this interest has lead to the demonstration that the
standard model of quantum computation can simulate efficiently the
dynamics of a few quantized chaotic models
\cite{Shack1998,GS2001,BCMS2001}; unfortunately, none of these
proposals indicate how to circumvent the measurement problem mentioned
above.

Recent work by Emerson {\it et al.} \cite{ewlc2002} proposes to study
statistical properties of the system's eigenvectors relative to a
perturbation as a signature of chaos. They also provide an efficient
procedure to measure these statistics using the standard model of
quantum computation. Their motivation for this work was to test the
validity of the signature. Indeed, it is not clearly established that
this signature is universal, perturbation independent and, most
importantly that the decay time does not scale with the size of the
system. 

Here, we concentrate on a different model (presumably weaker):
deterministic quantum computation with a single pseudo pure bit (DQC1)
which was introduced in \cite{KL1998}. In this setting, the initial
state of the $K+1$ qubits computer is $\rho =
\left\{\frac{1-\epsilon}{2} \II + \epsilon\kb{0}{0}\right\} \otimes
\frac{1}{2^K}\II$ where $0<\epsilon\leq 1$ is a constant. Note that,
from a computational complexity point of view, this is equivalent to a
model where the state of the first qubit is pure while the other ones
are completely random; we shall therefore assume that $\epsilon =1$ in
the remaining of the paper. The final answer is given by a finite
accuracy evaluation of the average value of $\sigma_z$ on the first
qubit. As for the dynamics, we assume that we are gifted with the
ability to exert coherent control over one and two qubits at a time.
This model is of particular interest since it is weaker than the
computational model offered by liquid state nuclear magnetic resonance
(NMR) quantum computing \cite{Cory2000}.  Such a computing device, we
shall show, can test for integrability using $O(N)$ physical
resources, given that the dynamics of the system of interest is
efficiently simulatable on the standard model of quantum computation
without ancillary pure qubits (or, more precisely, with no more than
$O(\log K)$ ancillary pure qubits).

In order to do so, we must first relate the theory underlying the
spectral property at the heart of our study; this is done in
Sec.~\ref{theory}.  We then show how it can be evaluated with $O(
N)$ physical resources in the DQC1 model. In Sec.~\ref{results}, we
present numerical results for canonical random matrices as well as for
a physical map, the nonlinear kicked top. Finally, we conclude with a
summary of our results and discussion of possible extensions.

\section{Level distribution}
\label{theory}

In the theory of quantum chaos, a key role is played by the statistics
of eigenvalues \cite{Haake2001,Stockmann1999}. In the case of systems
with a periodically time varying Hamiltonian the central dynamical
object is the Floquet operator, $\hat{F} = \tilde T[\exp\{-i\int_0^T
H(t)dt\}]$, that maps the state from one time to a time exactly one
modulation period $T$ later, $\tilde T$ is the time ordering operator.
The eigenvalues of $\hat{F}$ lie on the unit circle and may be
parameterized in terms of eigenphases, or quasi-energies, as
$\hat{F}|\phi_j\rangle=e^{-i\phi_j}|\phi_j\rangle$.

The random matrix conjecture asserts that the statistics of
eigenvalues of chaotic systems (dynamical systems and maps) is
typically well modeled by the statistics of the eigenvalues of random
matrices (hermitian Hamiltonians and unitary Floquet operators) with
the appropriate symmetries \cite{Haake2001,Stockmann1999}. While many
important mathematical results underpin the conjecture, a rigorous
proof is lacking and support rests on a very large accumulation of
numerical results.

An integrable system, by definition, possesses as many symmetries ---
constant of motion --- as degrees of freedom. One can thus write the
system's Hamiltonian as the direct sum of independent Hamiltonians
acting on smaller subspaces; one for each values of the constants of
motion. Some spectral properties of these Hamiltonians can thus be
reproduced by those of matrices that are the direct sum of independent
random Hermitian operators. The distribution characterizing the entire
spectrum is therefore given by the superposition of many independent
spectra; as a consequence, the correlations between levels vanish.
Thus, one might expect that the nearest neighbors level spacing
distribution (LSD) follow a Poisson law $prob(\phi_{j+1}-\phi_j = S) =
P(S) \sim e^{-\Gamma S}$, a straightforward consequence of their
statistical independence. This is indeed observed experimentally,
numerically, and most importantly can be derived formally
\cite{BT1977}.

On the other hand, chaotic systems possess no or just a few
symmetries. It can be shown \cite{Haake2001} that the LSD --- aside
from the systematic degeneracy following the symmetries --- obeys a
power law $P(S) \sim S^\beta e^{-\alpha S^2}$. The parameter $\beta$
characterizes the symmetries of the system; it is equal to $1$ when
the system possesses a time reversal symmetry and some geometric
invariance, $2$ when it has no symmetries, and $4$ when it has a time
reversal symmetry with Kramer's degeneracy. Similarly, we will refer to
the Poisson ensemble --- the characteristic ensemble of integrable
systems --- as $\beta = 0$.

The exact form of the LSD is not relevant to us; we shall capitalize
on the crucial distinct behavior of $P(S\rightarrow 0)$ for chaotic
and regular systems. In the former case, $P(S)$ reaches a minimum at
$S=0$: the levels tend to repel each other. In the latter case, $P(S)$
is maximal at $S=0$, a consequence of the levels statistical
independence called clustering.

With these considerations, one can predict the behavior of the ensemble average
form factors 
\begin{equation}
T_n = \left|Tr\{\hat F^n\}\right|^2 = \left|\sum_{j=1}^N
e^{-in\phi_j}\right|^2 
\end{equation} 
from which most spectral properties can be extracted.  For regular
systems, $Tr\{\hat F\} = \sum_j e^{-i\phi_j}$ behaves like the end
point of a random walk in the complex plane: each step having unit
length and uncorrelated random orientation $\phi_j$.  After $N$ steps,
the average distance from the origin is expected to be $\sqrt N$ so we
should find $\overline{T_1} = N$. For times $n>1$, the analysis is
identical; if the angles $\{\phi_j\}$ are statistically independent,
then so are $\{\phi_j^{(n)} = n\phi_j \mathsf{mod} (2\pi)\}$, $n$
taking positive integer values. We conclude that the ensemble average
form factors of integrable systems should be time independent and
equal to $N$.

For chaotic systems, more elaborate calculations are required for the
ensemble average form factors. They can be found in \cite{Haake2001};
here we shall simply give an approximate result for $0<n<N$ (accuracy
of order $10^{-2}$) known as the Wigner surmises:
\begin{equation}
\overline{T_n} =\left\{
\begin{array}{lc}
2n - n\sum_{m=1}^n \frac{1}{m+(N+1)/2} & \mathrm{for}~ \beta = 1 \\ 
n                                      & \mathrm{for}~ \beta = 2 \\
n + \frac{n}{2} \sum_{m=1}^n \frac{1}{N+1/2-m} & \mathrm{for}~ \beta = 4
\end{array}
\right. .
\label{Wigner}
\end{equation} 
Although simple arguments could not have indicated the exact behavior
of these form factors, we could have guessed their general form: they
are initially very small $\overline{T_1} \ll N$, and, as $n$ grows,
they reach the same value as the Poisson ensemble. Here, $Tr\{\hat
F\}$ is analog to a {\it anti-correlated} random walk in the complex
plane composed of $N$ unit steps. As a consequence of level repulsion,
each steps tend to be oriented in different directions; the
probability of finding two steps oriented within an angle $\epsilon$
decreases as $\epsilon^{\beta +1}$. Thus, the distance from the origin
after $N$ of these anti-correlated steps should definitely be smaller
than $\sqrt N$ which is the expected value for uncorrelated steps. As
$n$ grows, the phases $ n\phi_j \mathsf{mod} (2\pi)$ get wrapped
around the unit circle; the effect is analogue to superposing $n$
independent spectral distributions, blurring out the correlations.
When $n \sim N$, one should thus expect a behavior similar to the
Poisson ensemble.

It should be noted that the few symmetries of a chaotic system may
slightly affect the predictions of Eq.~\ref{Wigner}. The average
$\overline{T_n}$ were evaluated for fixed values of the constant of
motion. In what follows, we shall often neglect this point for sake of
simplicity. Nevertheless as long as the number of invariant subspaces
is small ($\ll \sqrt N$) this omission will not affect our
conclusions. For example, if a chaotic system possesses a symmetry
which breaks its Hilbert space into $k$ equal invariant subspaces, the
small $n$ behavior $\overline{T_n} \simeq n$ will be transformed into
$\overline{T_n} \simeq k^2n \ll N$, which is all that really matters
to us. One can circumvent this issue when some exact symmetries of
the system are known: it suffices to simulate the dynamics of the
system within an invariant subspace.

In the light of this analysis, it may seem that form factors
constitute a powerful tool to distinguish between classically regular
and chaotic systems. In particular, $T_n$ should clearly identify each
regime for small values of $n$. Nevertheless, the form factor $T_n$ of
a fixed Floquet operator $\hat F$ will generally fluctuate about the
ensemble average $\overline{T_n}$. Thus, we seek a signature of an
{\it ensemble property} on a {\it single element} drawn from this ensemble.

The solution is to use a version of the ergodic theorem. If we
normalize out the explicit time dependence of the form factors, an
average over a time interval $\Delta n$ reproduces the effect of an
ensemble average. More precisely, one can show \cite{Haake2001} that
\begin{equation}
\left\langle T_n / \overline{T_n}\right\rangle
=\frac{1}{\Delta n} \sum_{n' = n-\frac{\Delta n}{2}}
^{n+\frac{\Delta n}{2}}{T_{n'}/\overline{T_{n'}}}
\label{form_av}
\end{equation}
converges to $1$ with a variance $\sigma^2$ bounded by $1/\Delta n$.
For large $N$, we can thus use the first $\Delta n \ll N$ form factors
to determine whether the Floquet operator belongs to a polynomial or
Poisson ensemble. Since the value of $\overline{T_n}$ --- hence the
matrix ensemble --- are needed to compute Eq.~\ref{form_av}, we shall
proceed by hypothesis testing: for which choice of $\overline{T_n}$
($\overline{T_n} = N$ regular, $\overline{T_n} \simeq n$ chaotic) does
Eq.~\ref{form_av} converge to 1? In other words, we need to determine
which of the two variables
\begin{equation}
t_0 = \frac{1}{\Delta n}\sum_{n=1}^{\Delta n} \frac{T_n}{N}
\mathrm{\ \ or\ \ }
t_1 = \frac{1}{\Delta n}\sum_{n=1}^{\Delta n} \frac{T_n}{n}
\label{hyp}
\end{equation}
is most probably drawn from a distribution centered at $1$ with
$1/\sqrt{\Delta n}$ standard deviation. If we restrict our attention
to a regime where $\Delta n \ll N$, both hypothesis cannot have high
probabilities simultaneously \footnote{It is important to note that $\Delta n$ needs not to increase with $N$, in fact it is quite the opposite. The probability of error scales like the overlap of two Gaussian distributions of width $\sigma = 1/\Delta n$ and centered about points $\mu_1$ and $\mu_2 \simeq N\mu_1$: clearly, this overlap decreases with $N$.}. On the other hand, when the
probabilities of both hypothesis are low, the test is inconclusive.
Nevertheless, remember that the presence of symmetries in a chaotic
system shifts the value of the distribution by a factor $k^2$ where
$k$ is the number of invariant subspaces. For $k^2 \ll N$, this should
be clearly distinguishable from the value of a regular system. This
should not be seen as a bug but a feature of our approach allowing one
to estimate $k$, the number of invariant subspaces. 

Applying this test to a particular dynamical system would require one to
compute the spectrum of the Floquet operator. If one were to
try and simulate a dynamical map on a quantum computer with $K$
qubits, a direct computation would require determining all
$N=2^K$ eigenvalues. In the next section we will construct a quantum
circuit that would enable the form factors themselves to be extracted
with $O(N)$ physical resources thus allowing a direct test of
non-integrability that circumvented the need to explicitly compute all
eigenvalues.

\section{Quantum algorithm}
\label{algo}

\begin{figure}[!htb]
\centering
\includegraphics[width=7cm]{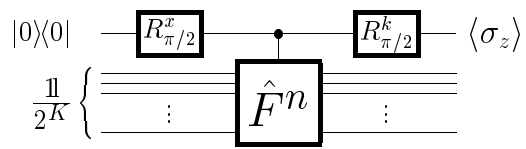}
\caption{Quantum circuit evaluating the trace of $\hat F^n$. The gates
  $R_\theta^k$ are rotation in the Bloch sphere by an angle $\theta$
  around axis $k=x$ or $y$. When $k$
  is set to $x$, we get the real part of the trace while $k=y$ yields
  the imaginary part.}
\label{circuit}
\end{figure}

The DQC1 algorithm evaluating the form factor is based on the idea
reported in \cite{MPSKLN2002} of using a quantum computer as a
spectrometer.  The circuit is shown at Fig.~\ref{circuit} where $K =
\lceil \log_2 N \rceil$.
By hypothesis, we are able to efficiently simulate the dynamics of the
system under study so the gate $\hat F^n$ only requires a polynomial
(in $n$ and $K$) number of elementary gates to be constructed. Here,
it is not $\hat F^n$ we wish to implement but a coherently controlled
version of it, i.e. a linear gate acting on $K+1$ qubits which applies
$\hat F^n$ to the last $K$ qubits when the first qubit is in state
$\ket 1$ and doesn't do anything when it is in state $\ket 0$. Given
the circuit for $\hat F^n$, standard techniques can be used to
construct a controlled version of it at polynomial cost
\cite{BBCDMSSSW1995}.

It should also be emphasized that the $K$ qubits on which the Floquet
operator is applied generate a Hilbert space of dimension $2^K$ which
might be larger than the simulated system's Hilbert space. Thus, when
applying $\hat F$ to those qubits, one really applies $\hat F \oplus
U$ where ideally $U$ is the identity operator on $2^K-N$ states; it
can be any other unitary operator as long as its trace can be
evaluated. The effect of these extra dimensions will be to add a
contribution $Tr\{U\}/N$ to the output signal which should be
systematically subtracted as we shall henceforth assume.

The output of this computation will be the real and imaginary part of
$(Tr\{\hat F^n\})/2^K$ when the last rotation is made about axis $k=x$
and $k=y$ respectively. Thus, our task is to distinguish between a
signal whose amplitude is of order $1/N$ (chaotic dynamics) and one of
order $1/\sqrt N$ (regular dynamics) which can be achieved using
$O(N)$ physical resources. In the special case of NMR quantum
computing, one can for example increase the size of the sample by a
factor $N$ as the size of the system increases, or simply repeat the 
procedure $N$ times and sum up the outputs. We thus get a
quadratic advantage over all known classical algorithms.

\section{Numerical results}
\label{results}

\subsection{Random Matrices}

\begin{figure}[!htb]
  \centering 
\includegraphics[height=4.5cm]{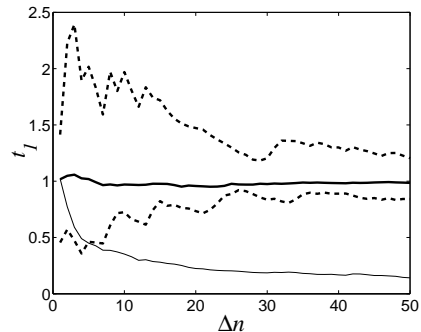}
\caption{The two dash lines show $t_1$ as a function of $\Delta n$ (Eq.~\ref{hyp}) for two random unitary matrices drawn from the ensemble $\beta =2$. The heavy full line is the value of $t_1$ averaged over 50 such random matrices while the light line shows its variance, which drops as $1/\sqrt{\Delta n}$ as expected.}
\label{rand_matrices}
\end{figure}

Before applying our general proposal to a physical model, we give a
numerical example illustrating the main results used from random
matrix theory: the ergodic theorem of Eq.~\ref{hyp}. In order to
estimate the average and variance of $t_0$ and $t_1$ in a given
universal matrix ensemble, we draw many random matrices $U^{(k)}$ from
the ensemble and numerically evaluate each quantity. As an example,
we have generated 50 random matrices from the $\beta = 2$ ensemble ---
the set of unitary matrices with no symmetries.  This is illustrated
on Fig.~\ref{rand_matrices}, where the matrices are of size $600
\times 600$. For each random matrix $U^{(k)}$ drawn from this
ensemble, we can compute $t_1(U^{(k)})$ as functions of $\Delta n$.
Two such curves (dashed) are plotted on Fig.~\ref{rand_matrices}.  By
applying this procedure to many samples (here 50), we can estimate the
average of $t_1$ and its fluctuations:
\begin{equation}
\langle t_1 \rangle = \frac{1}{50}\sum_{k=1}^{50}
t_1(U^{(k)}),\ \
\langle (t_1)^2 \rangle = \frac{1}{50}\sum_{k=1}^{50}
[t_1(U^{(k)})]^2.
\end{equation}
The average $\langle t_1 \rangle$ and mean deviation $\sigma =
\sqrt{\langle (t_1)^2 \rangle - \langle t_1 \rangle^2}$ are also
plotted on Fig.~\ref{rand_matrices} (heavy and light full line
respectively): as expected, $t_1$ converges to 1 as $1/\sqrt{\Delta
  n}$. The same procedure can be applied to $t_0$; nevertheless, since
$t_1$ does converge to 1 in this ensemble, $t_0$ obviously does not
since it differs by a factor of roughly $N/\Delta n \simeq 20$ for the
range of $\Delta n$ we have studied. Of course, this difference would
vanish when $\Delta n$ approaches $N$ since the form factor of any
universal ensemble converge to those of the Poisson ensemble (see
Sec.~\ref{theory}); this is why we must restrict our study to $\Delta
n \ll N$. The same conclusions can be reached for the other ensembles
characterizing chaotic systems, i.e. $\beta = $1, 2, and 4.

Similarly, had the matrices $U^{(k)}$ been drawn from the $\beta = 0$
ensemble --- the set of matrices characterizing regular systems --- we
would have observed $t_0$ converging to $1$ as $1/\sqrt{\Delta n}$
while $t_1$, smaller by a factor of roughly $\Delta n/N$, would
roughly vanished. From these considerations, the hypothesis test
``$t_0$ converges to 1'' versus ``$t_1$ converges to 1'' allows us to
discriminate between random matrices drawn from $\beta=0$ and those
drawn from one of the $\beta=$ 1, 2, or 3, with a probability of error
decreasing as $1/\sqrt{\Delta n}$. Thus, as long as the random matrix
conjecture holds, it should also allow to discriminate between regular
and chaotic motion.

\subsection{Kicked Top}

We now focus our attention on a physical model of great interest for its good
agreement with random matrix theory: the nonlinear kicked top. We
write the Floquet operator in its most general form following Haake
\cite{Haake2001}, $\hat F = U_zU_yU_x$ with
\begin{equation}
U_k = \exp\left\{ -i\frac{\tau_k J_k^2}{2j+1} -i\alpha_kJ_k\right\}
\label{floquet}
\end{equation}
where the $J_k$, $k = x,\ y,$ and $z$, are the canonical angular
momentum operators. We conveniently define a parameter vector $\mathbf
p = (\alpha_x,\alpha_y,\alpha_z,\tau_x,\tau_y,\tau_z)$. Some authors
use a restricted form of this Floquet operator where only $\tau_z$ and
$\alpha_y$ are non-zero. Since $[\hat F,\mathbf{J}^2] = 0$, the value
of the angular momentum $j$ --- which appears in Eq.~\ref{floquet} ---
is conserved. The dimension of the Hilbert space is simply
given by $N = 2j+1$.

\begin{figure}[!htb]
\centering
\includegraphics[height=4.5cm]{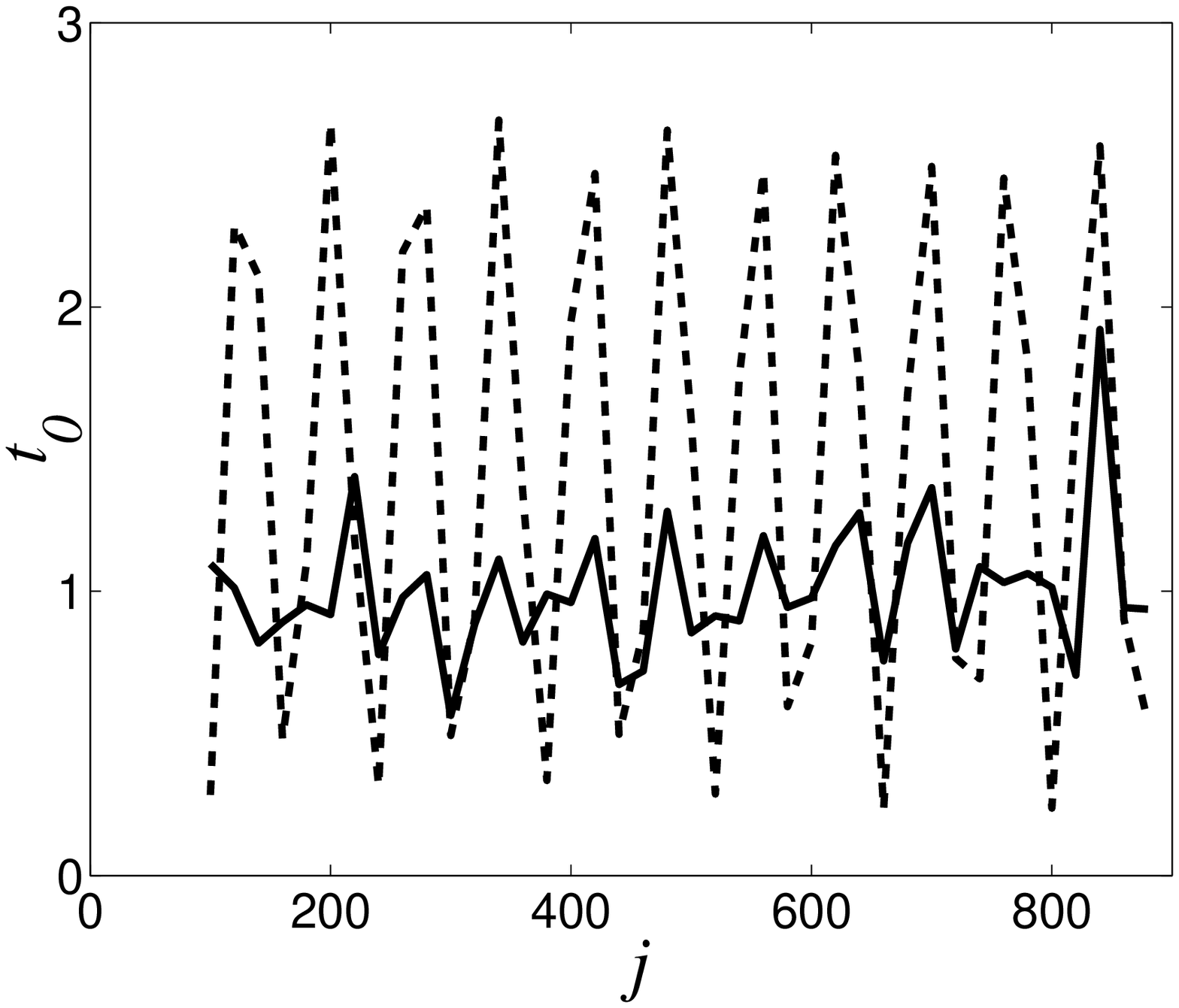}
\caption{Value of $t_0$ (Eq.~\ref{hyp}) of the kicked top in a regular regime $\mathbf{p}_r = (0,0,1,0,0,10)$ as in \cite{Haake2001} for different values of $j$. Dashed curve: $\Delta n=1$ so it is simply $Tr\{\hat  F\}/N$. Full curve: To decrease the fluctuation, we have used the ergodic averaging over the first $\Delta n = 30$ normalized form factors $Tr\{\hat F^n\}/N$, $n=1, 2,\ldots 30$.}
\label{regular}
\end{figure}

\begin{figure}[!htb]
\centering
\includegraphics[height=4.5cm]{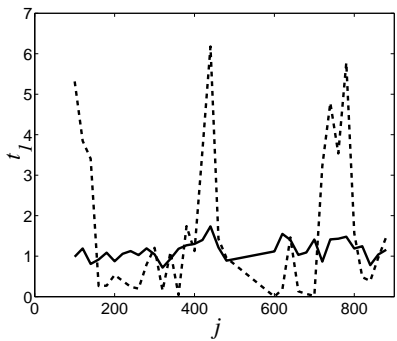}
\caption{Value of $t_1$ (Eq.~\ref{hyp}) of the kicked top in a chaotic regime $\mathbf{p}_c = (1.1,1,1,4,0,10)$ as in \cite{Haake2001} for different values of $j$. Dashed curve: $\Delta n=1$ so it is simply $Tr\{\hat  F\}$. Full curve: To decrease the fluctuation, we have used the ergodic averaging over the first $\Delta n = 30$ normalized form factors $Tr\{\hat F^n\}/n$, $n=1, 2,\ldots 30$.}
\label{chaotic}
\end{figure}

By adequately choosing the parameters $\mathbf p$, the kicked top can
be either in a regular or chaotic regime, see \cite{Haake2001} for
more details.  Thus, we can evaluate $t_0$ and $t_1$ of Eq.~\ref{hyp}
in both regimes and verify that they indeed allow to
discriminate between them.  This is presented on Figs.~\ref{regular}
and \ref{chaotic} for different values of the total angular momentum
$j$. On Fig.~\ref{regular}, the system is in a regular regime; we have
only plotted $t_0$ since $t_1$ is larger by a factor proportional to
$j$ so clearly does not converge to 1. Similarly, only the value of
$t_1$ is exhibited of Fig.~\ref{chaotic}. Notice that while the
ergodic averaging decreases the fluctuations, it is not essential to
discriminate between regular and chaotic. Indeed, the scale of the
fluctuation is extremely small compared to $j$ which is the factor by
which $t_0$ and $t_1$ differ.

\begin{figure}[!htb]
\centering
\includegraphics[height=4.5cm]{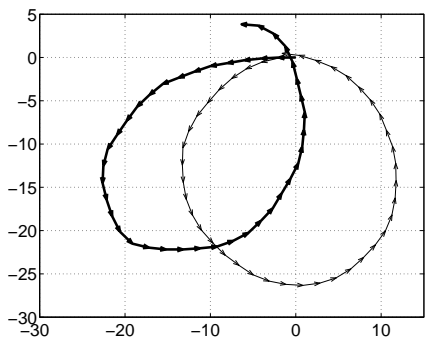}
\caption{Vectorial representation of eigenphases: $\sum_j
  (\cos\phi_j,\sin\phi_j)$ where the $\phi_j$ have been ordered in
  increasing order. The $\alpha$ and $\tau$ parameters of the Floquet
  operator Eq.~\ref{floquet} are tuned so the system is in a regular
  regime (heavy vectors) and a chaotic regime (light vectors) as in
  Figs.~\ref{regular} and \ref{chaotic}. The value of $j$ is $20$ so
  each curve contains $41$ vectors.}
\label{walk}
\end{figure}

The analogy with a random walk in the plane can also be illustrated
graphically. On Fig.~\ref{walk} we have plotted the sum of the
eigenvalues vectorially. The apparent structure of the vectors is
purely artificial, the eigenphases were ordered in increasing order
(the sum of vectors is obviously a commutative operation); we have
chosen this ordering to facilitate the presentation.

The effect of LSD are striking on Fig.~\ref{walk}. The light vectors
(chaotic regime) are arranged in an almost perfect circle; eigenphases
tend to be equally separated. On the other hand, the heavy vectors
(regular regime) are quite often aligned in an almost straight line;
a manifestation of level clustering. As a consequence, the heavy vectors
end up further apart from the origin than do the light vectors; on
average, these distances differ by a factor $\sqrt N$. 

\begin{figure}[!hbt]
\centering
\includegraphics[height=4.5cm]{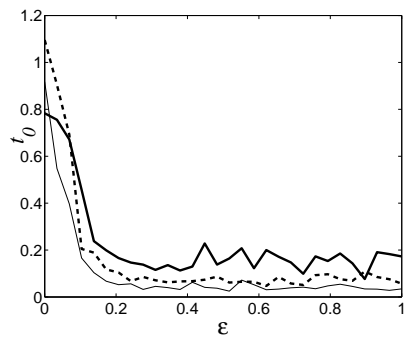}
\caption{Value of $t_0$ (Eq.~\ref{hyp}) with $\Delta n=30$ for different system size: Full heavy line $j=50$; Dashed line $j=100$; Light full line $j=200$. }
\label{trans}
\end{figure}

Finally, we can use the form factor to study the transition between
regular and chaotic motion. To do so, we let the parameter vector
continuously vary from its regular value to its chaotic value:
$\mathbf{p} = (1-\epsilon)\mathbf{p}_r + \epsilon\mathbf{p}_c$ (see
figure captions \ref{regular} and \ref{chaotic}). For $\epsilon = 0$,
the expected value of $t_0$ is 1.  As $\epsilon$ increases, the system
enters a chaotic regime; when chaos has fully developed, $t_0$ should
vanish as $1/N$.  This is indeed observed on Fig.~\ref{trans}, where
we have plotted $t_0$ as a function of $\epsilon$ for different system
sizes. Moreover, the results indicate that the transition to chaos
becomes more sensible as the size of the system increases.

\section{Conclusion}
\label{conclusion}

We have shown that, using a single bit of quantum information, we can
test whether the spectrum of a unitary matrix obeys a Poisson or
polynomial law. Under the random matrix conjecture, this can be used
to determine whether the system has a regular or chaotic behavior in
its classical limit. The idea relies on estimating the averaged form
factor using the ergodic theorem which roughly states that a time
average can reproduce an ensemble average. The form factors in a
regular and chaotic regime differ by a factor of $\sqrt{N}$ and the
output signal of our computation decreases as $1/N$: the required
physical resources thus scale as $N$. This is a quadratic
improvement over all known classical algorithms. We are currently
investigating a different signature of quantum chaos which might not
suffer from this signal loss and hence, could offer an exponential
speed up.

This result gives a new insight on the nature of the potential
computational speed up offered by quantum mechanics. In particular, it
provides a strong argument towards the computational power of mixed
states quantum computing.

We acknowledge Harold Ollivier for stimulating discussions and careful
reading of this manuscript, and Howard Wiseman for pointing out an error 
in an earlier version of this work. DP is financially supported by Canada's NSERC, and RL by NSERC and CIAR.

\end{document}